\documentclass{egpubl}
\usepackage{VMV2024}

\ArxivPaper

\usepackage[T1]{fontenc}
\usepackage{dfadobe}  

\usepackage{cite}  
\BibtexOrBiblatex
\electronicVersion
\PrintedOrElectronic
\ifpdf \usepackage[pdftex]{graphicx} \pdfcompresslevel=9
\else \usepackage[dvips]{graphicx} \fi

\usepackage{egweblnk} 

\title[Application of 3DGS for Cinematic Anatomy]{Application of 3D Gaussian Splatting \\
for Cinematic Anatomy on Consumer Class Devices}

\author[S. Niedermayr\& C. Neuhauser\& K. Petkov \& K. Engel \& R. Westermann]{
    \parbox{\textwidth}{\centering
    S. Niedermayr$^1$\orcid{0009-0008-3370-0149}, C. Neuhauser$^1$\orcid{0000-0002-0290-1991}, K. Petkov$^2$\orcid{0009-0008-1914-4625}, K. Engel$^2$\orcid{0009-0001-1423-898X}, R. Westermann$^1$\orcid{0000-0002-3394-0731}
    }\\
    \parbox{\textwidth}{\centering
    $^1$Technical University of Munich, $^2$Siemens Healthineers
    }
}

\usepackage{xspace}

\makeatletter
\DeclareRobustCommand\onedot{\futurelet\@let@token\@onedot}
\def\@onedot{\ifx\@let@token.\else.\null\fi\xspace}

\def\etal{\emph{et al}\onedot}

\usepackage{amsmath}
\usepackage{amssymb}
\usepackage{siunitx}
\usepackage{multirow}
\usepackage{algorithm}
\usepackage{algpseudocode}
\usepackage[capitalise]{cleveref}
\usepackage[utf8]{inputenc}
\usepackage{tabu}
\usepackage{booktabs}
\usepackage{subcaption}
\usepackage{todonotes}

\begin{document}

\teaser{
 \includegraphics[width=\textwidth]{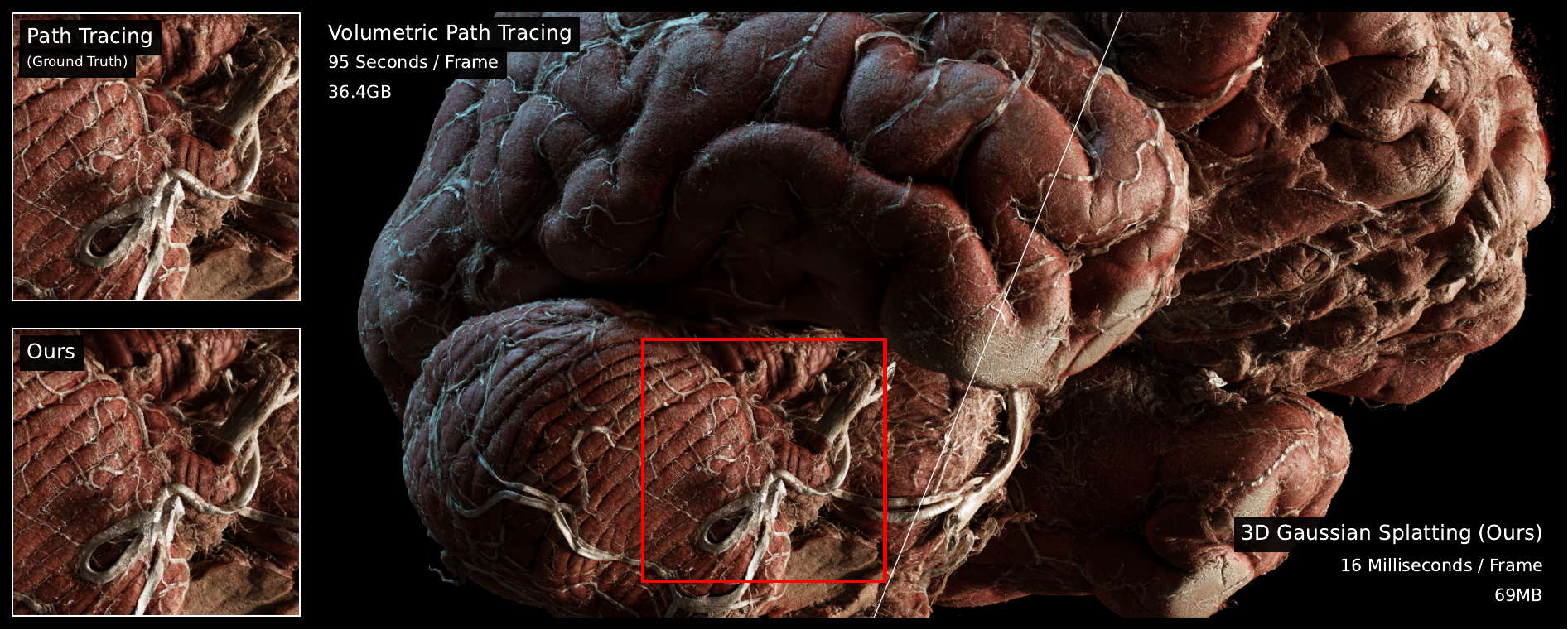}
 \centering
  \caption{Left: Path-traced image of a 36 GB HiP-CT data set, rendered in 95 seconds on a high-end GPU. Right: Gaussian splat representation of the same data set requiring 69 MB and rendered at 60 frames per second. The pixel resolution is 2048x2048. }
\label{fig:teaser}
}

\maketitle
\begin{abstract}
Interactive photorealistic rendering of 3D anatomy is used in medical education to explain the structure of the human body.
It is currently restricted to frontal teaching scenarios, where even 
with a powerful GPU and high-speed access to a large storage device where the data set is hosted, interactive demonstrations can hardly be achieved.
We present the use of novel view synthesis via compressed 3D Gaussian Splatting (3DGS) to overcome this restriction, and to even enable students to perform cinematic anatomy on lightweight and mobile devices. 
Our proposed pipeline first finds a set of camera poses that captures all potentially seen structures in the data. High-quality images are then generated with path tracing and converted 
into a compact 3DGS representation, consuming < 70 MB even for data sets of multiple GBs. This allows for real-time photorealistic novel view synthesis that recovers structures up to the voxel resolution and is almost indistinguishable from the path-traced images.

\ccsdesc[300]{Computing methodologies~Computer graphics; Rendering}

\printccsdesc   
\end{abstract}  
\section{Introduction}

\maketitle

Cinematic Anatomy (CA) is an immersive anatomy learning application, 
which is designed to improve anatomy education through the use of photorealistic 3D 
rendering via path-tracing\cite{CVR01}. Instead of real 3D anatomy models, it utilizes volume data provided by medical scanning devices.
The application is used in the field of anatomy education, e.g., in the JKU medSPACE~\cite{jku-medspace}, a lecture space for teaching anatomy. It is used for teaching the diverse and complex individual human anatomy, anatomical variations, and pathology, to enhance learners' competency with immersive photorealistic 3D visualization of 
data from real patients. 
Several studies have shown the benefits of using photo-realistic volumetric rendering of clinical volume data for teaching and understanding anatomy~\cite{GLEMSER2018e283, BINDER2019159, STEFFEN2022151905}. 

Additionally to stereoscopic projection modes for frontal education, students need to run CA on their mobile devices 
for personalized learning experiences. In addition to significant performance losses on such devices, the portability of the created content 
is however often limited by the data size, especially when employing data from high-resolution imaging modalities like photon-counting CT, 7 Tesla MRI and phase-contrast CT~\cite{Walsh2021}. 
Thus, CA is mostly used in frontal teaching scenarios, where the demonstrator uses a powerful GPU and has high-speed access to a large storage device where the data set is stored. As \cref{fig:teaser} demonstrates, even then is it difficult to render the data at interactive rates.    


We demonstrate that differentiable 3DGS~\cite{kerbl3Dgaussians}, which reconstructs a 3D Gaussian scene representation 
from images of this scene, can address the limitations of CA. The Gaussian representation can be rendered at high speed
from arbitrary views, 
avoiding time-consuming path tracing. 
In combination with compressed 3DGS~\cite{niedermayr2024compressed}, the memory consumption of the Gaussian representation is significantly reduced, and with GPU rasterization, 3D Gaussian splatting runs efficiently even on mobile devices.
\cref{fig:teaser} demonstrates these properties with a high-resolution CT scan. 



\textbf{Contribution.} To use compressed 3DGS for CA, we propose a processing pipeline including the following adaptations: 

\begin{itemize}
    \item We extend the view selection method proposed by Kopanas and Drettakis~\cite{kopanas2023improving} for volume rendering to automatically find a set of cameras that captures all potentially seen structures under the current transfer function setting.
    \item We extend 3DGS with differentiable alpha channel rendering to create background-free reconstructions and drastically improve the reconstruction of translucent materials.
\end{itemize}

We analyze the quality, performance, and memory requirements with several high-resolution data sets. Training images are rendered with a publically available CA tool.
The results demonstrate that the memory requirement is significantly below the initial data size. 
Since the renderable representation is so small, students can quickly download it over low-bandwidth channels and render on their mobile devices. 
Rendering performance is about two orders of magnitudes faster than optimized path tracing, with almost no perceptible loss of image quality. 

\noindent{\textbf{Limitations.}} The use of 3DGS for CA comes with the following limitations: Firstly, lighting conditions are baked into the 3D Gaussian representation and cannot be changed during rendering. Secondly, due to the use of preset transfer functions and clip planes, the approach is less effective in supporting interactive volume exploration. Overcoming these limitations is difficult, and we discuss possible improvement strategies at the end of our work. 

\begin{figure*}[ht]
    \centering
    \includegraphics[width=\linewidth]{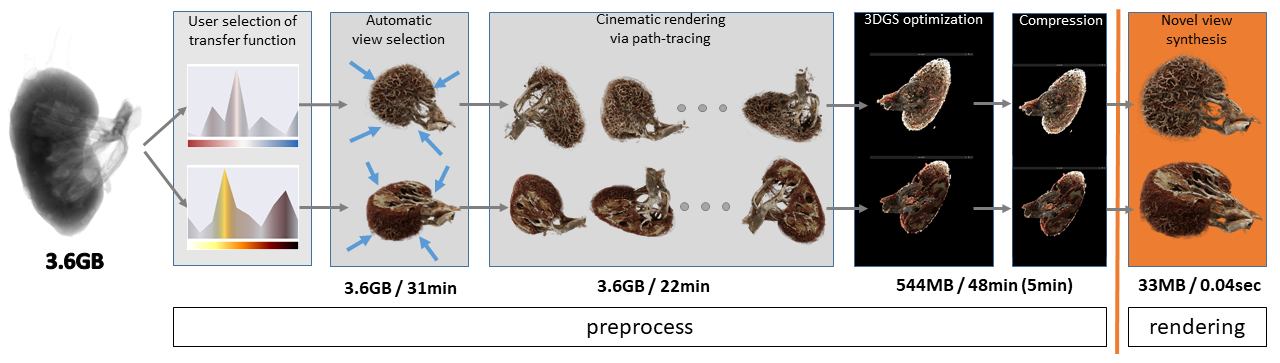}
    \caption{CA pipeline for a $1510 \times 1706 \times 1415$ HiP-CT data set requiring 3.6 GB of memory. Numbers below each stage indicate the required memory at this stage and its computation times. 3D Gaussian splatting (3DGS) optimization uses 99 path traced images, and first generates the raw Gaussian representation in 48 minutes before it is compressed to 33 MB in 5 minutes.}
    \label{fig:overview}
\end{figure*}

\section{Related Work}
3DGS \cite{kerbl3Dgaussians} builds upon elliptical weighted average (EWA) volume splatting \cite{zwicker_ewa_2001} to efficiently compute the projections of 3D Gaussian kernels onto the 2D image plane. In addition, the number and parameters of the Gaussian kernels that are used to model the scene are optimized with differentiable rendering. 
Mip-Splatting~\cite{Yu2023MipSplatting} modifies 3DGS by integrating anti-aliasing with a 3D smoothing and a 2D Mip filter. It achieves improved quality of novel views at scales the Gaussian representation has not been optimized for. A number of approaches have concurrently proposed to convert the 3D Gaussian representations generated by 3DGS into a more compact form \cite{niedermayr2024compressed,lee2024compact}. For typical scenes, the memory requirements of 3DGS is below 50 MB without any noticeable differences in the reconstructed images. 

3DGS for novel view synthesis overcomes in particular the difficulties of voxel-based approaches \cite{mildenhall2020nerf,Fridovich-Keil_2022_CVPR} to deal with sparsity. Even though adaptive hash grids \cite{mueller2022instant}, tensor decomposition \cite{Chen2022ECCV} or variants using dedicated compression schemes \cite{Li_2023_CVPR,Rho_2023_CVPR} can effectively reduce the required memory, they use volume ray-casting and, thus, require high-end GPUs to achieve reasonable rendering performance. The same limitation holds for differentiable volume rendering \cite{diff_fast_2022}, which, similar in spirit to 3DGS, optimizes optical properties on a dense voxel grid using image-based loss functions.

3DGS optimizes a 3D scene representation from photorealistic images of that scene, which are generated with volumetric path-tracing  \cite{pharr2023physically,novak2014residual,novak2018monte,NimierDavid2020Radiative}. A number of approaches have  
previously attempted to improve the performance of path tracing via image denoising~\cite{NeuralDenoisingMed,NeuralDenoisingVolRen,RealtimeDenoisingVPT}, photon mapping~\cite{CVRPhotonMapping}, illumination caching~\cite{VolumeSurfaceIlluminationCaching} and adaptive temporal sampling~\cite{AdativeTempSamplingVPTMed}. Despite achieving remarkable performance gains at high quality, all these approaches require a rendering system with enormous memory resources to host high-resolution data sets as well as huge computational power to perform ray tracing with such data. It is fair to say that high-quality path-tracing on consumer class hardware is impossible today. 

In principle, the memory requirements of CA can be addressed with Scene Representation Networks (SRNs) ~\cite{mescheder2019occupancy,chen2019learning,park2019deepsdf}, i.e., fully-connected neural networks that learn to encode a surface model as an implicit 3D function. Lu~\etal~\cite{lu2021compressive} demonstrate the use of SRNs for volume data compression, by overfitting a network to a volume data set. 
This approach, however, comes at the expense of subsequent network evaluations during rendering, which makes even GPU-friendly ray-marching implementations~\cite{weiss_fast_2022} significantly slower than 3DGS. 

A challenging problem in novel view synthesis is to find an as small as possible set of camera poses for generating the training and test images. Note that this problem is different to the problem of viewpoint optimization in visualization, where visualization parameters are optimized to find a single best viewpoint, e.g., by using entropy-based ~\cite{ji2006dynamic,vazquez2008representative,tao2009structure,chen2010information,weiss_fast_2022} or similarity-based~\cite{tao2016similarity,yang2019deep} loss functions that  guide an optimizer. 
For SRNs, Kopanas and Drettakis~\cite{kopanas2023improving} have introduced an algorithm to automatically optimize the placement of cameras so that improved coverage of a scene is achieved.
We adapt this algorithm to work with volumetric data sets. 



\section{Cinematic Anatomy Pipeline}
The different stages of the proposed CA pipeline are shown in \cref{fig:overview}.
After loading a data set, one or multiple so-called presets are selected by the user.
A preset includes the transfer function setting as well as material classifications and fixed clip planes that are used to reveal certain anatomical structures.

For each preset, multiple views capturing all potentially seen structures in the data are computed (cf.~\cref{sec:view-selection}).
In this way, we recover structures in the final object representation which are not seen when generating images with camera positions on a surrounding sphere.
These views are handed over to a physically-based renderer, i.e., a volumetric path tracer, which renders one image for every view using the corresponding preset (cf.~\cref{sec:image-generation}).

Once the images for a selected preset have been rendered, 3DGS generates a set of 3D Gaussian splats with shape and appearance attributes so that their rendering matches the given images.
Once the parameters of the Gaussians are computed via differentiable rendering, they are compressed using sensitivity-aware vector quantization and entropy encoding (cf.~\cref{sec:3dgs}).
The final compressed 3DGS representation is rendered with WebGPU using GPU sorting and rasterization of projected 2D splats, with a pixel shader that evaluates and blends the 2D projections in image space. We embed Mip-Splatting \cite{Yu2023MipSplatting} to account for different levels of detail and enable smooth transitions when the focal length is increased. 


\subsection{View Selection}\label{sec:view-selection}

Novel view synthesis requires that all visible scene parts are covered in the training images.
Kopanas and Drettakis~\cite{kopanas2023improving} propose an automatic camera placement algorithm for SRNs which maximises observation frequency and angular uniformity. The observation frequency lies between 0 (no camera observes a point) and 1 (all cameras observe a point). The angular uniformity considers the total variation distance between a 2D histogram of the directions of cameras observing a point in spherical coordinates and a uniform distribution. 
The algorithm iterates over batches of 1000 randomly sampled camera poses. In each iteration the cameras lying inside of or too close to occupied space are rejected. From the remaining camera poses the one is selected resulting in the highest improvement of the reconstruction, measured by observation frequency and angular uniformity. 

We propose two modifications of this algorithm for 3DGS of volumetric data sets.
Firstly, we observe that due to the high dimensionality of the search space, a huge number of cameras needs to be evaluated and optimal camera poses might be missed. To avoid this, we use Bayesian Optimal Sampling (BOS)~\cite{bayesopt_1989,bayesoptbook_2023} to adaptively place cameras in regions that are more promising to yield an improved maximum (called \textit{exploitation}) or in previously rather unexplored regions (called \textit{exploration}). In this way, the chance of missing optimal camera poses is significantly reduced, and fewer training images need to be generated. 
Secondly, instead of using a binary visibility indicating whether a point lies inside or outside the camera frustum, we use a continuous visibility that also considers (partial) occlusion. At each voxel, GPU ray marching is performed to compute the maximum transmittance, which is then used as a visibility indicator. For data sets not fitting into GPU memory, we perform this step with a lower-resolution copy.


BOS applies a probabilistic (usually Gaussian) surrogate model and an acquisition function.
The former expresses Bayesian belief about the output of the objective function derived from prior evaluations, and the latter is used for selecting the next set of parameters for evaluating the objective function.
For the acquisition function, the upper confidence bound~\cite{UCB} is used with parameter $\kappa = 10$, which controls the trade-off between exploitation and exploration.
The value was empirically determined to work well with all test data sets, but can also be subjected to hyperparameter optimization either regarding the energy term by Kopanas and Drettakis~\cite{kopanas2023improving} or the reconstruction quality with respect to images in a training set. We make use of the publically available software library Limbo~\cite{Limbo} to perform the optimization process.

\begin{figure}[t]
    \centering
    \begin{subfigure}[b]{0.45\linewidth}
        \includegraphics[width=\linewidth]{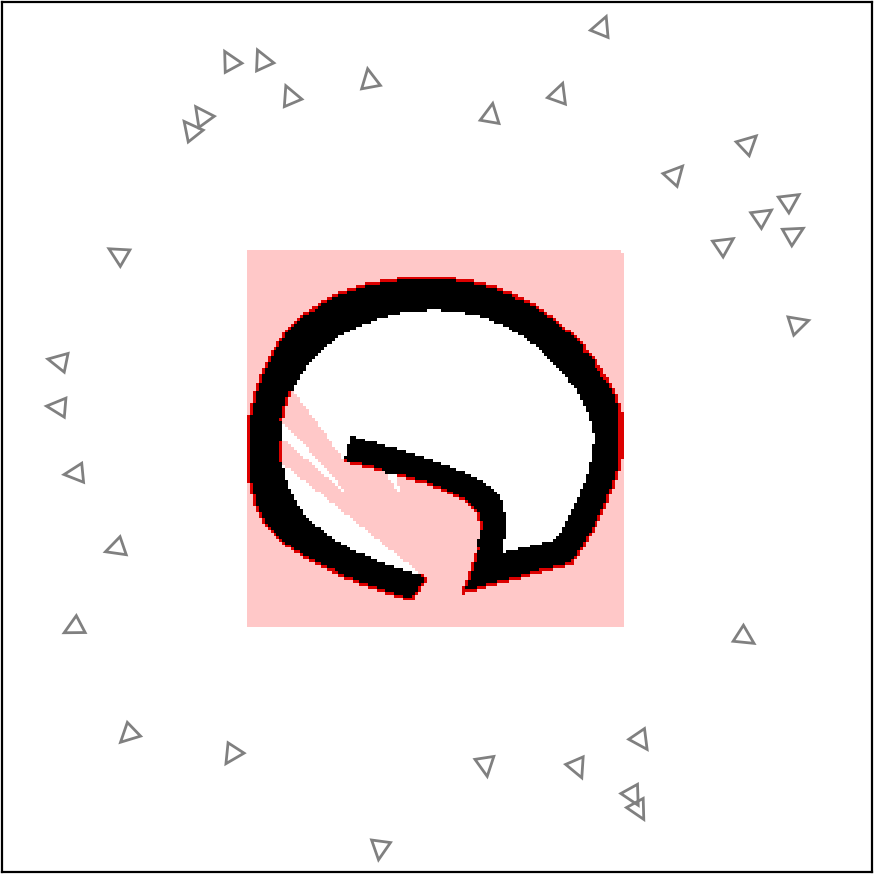}
        \caption{Pure hemisphere sampling.}
    \end{subfigure}
    \hfill
    \begin{subfigure}[b]{0.45\linewidth}
        \includegraphics[width=\linewidth]{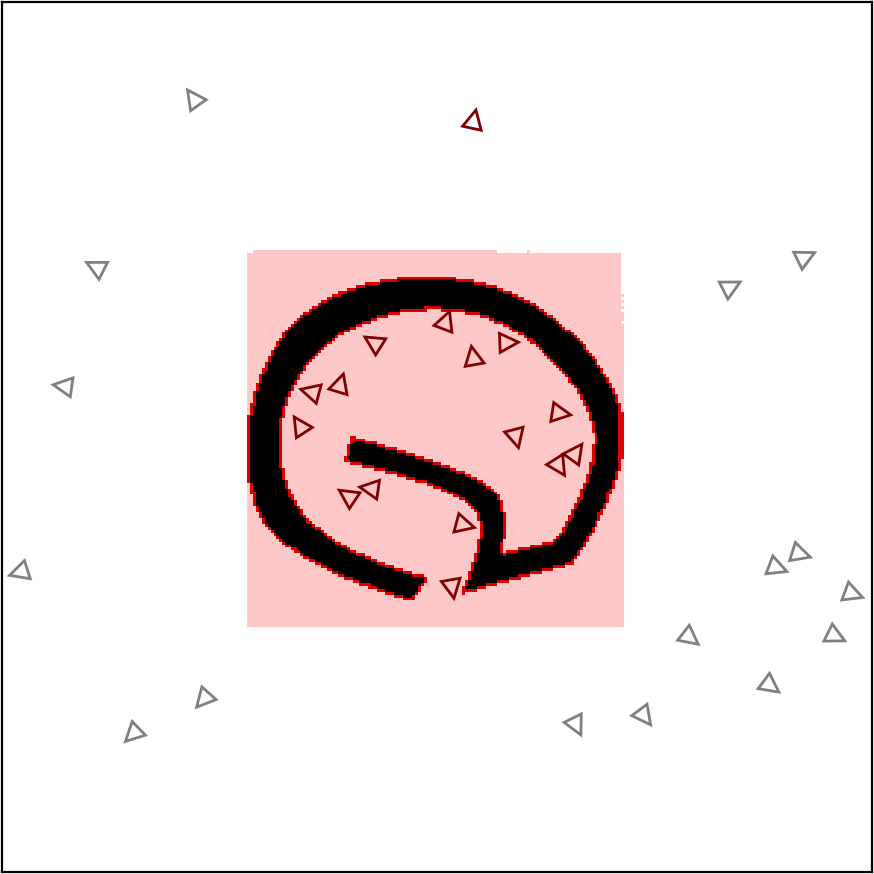}
        \caption{Bayesian optimal sampling.}
    \end{subfigure}
    \caption{View selection strategies for a concave 2D test data set (black). Left: Random sampling of camera poses on a bounding circle cannot capture interior structures. Right: With BOS, given a sparse set of initial poses (grey), additional poses (red) are automatically determined to capture otherwise not seen object parts. Red area indicates seen domain parts.}
    \label{fig:view-selection-2d}
\end{figure}

In \cref{fig:view-selection-2d}, we showcase the ability of automatic view selection using BOS for improved capturing of internal structures of a concave 2D test data set. When randomly sampling cameras on the hemisphere around an object, structures in the interior are missed. When sampling only a small set of cameras on the hemisphere, BOS selects few additional camera poses which capture insufficiently seen structures on the outside and not yet captured structures on the inside. We demonstrate the resulting improvements in the reconstruction quality with 3DGS in \cref{sec:eval-viewsel}. We further compare the convergence rate and performance of BOS and random sampling~\cite{kopanas2023improving} in the supplementary material.

\subsection{Image Generation}\label{sec:image-generation}
We render volume data with Monte Carlo volume path tracing from multiple views to generate a set of training images. 
Delta tracking~\cite{DeltaTrackingWoodcock} is used to determine 
a scattering event (the Henyey-Greenstein phase-function~\cite{HenyeyGreenstein} 
is applied to determine the next ray direction), an absorption event (the path is terminated and the emissive color is regarded as the path contribution) or a null collision (the path is followed unchanged).
A surface intersection is assumed when the density gradient magnitude exceeds a user-specified iso-value. 
Global illumination is then simulated by generating a reflection event for the surface with the new ray direction sampled proportional to the probability density function of a chosen reflectance distribution function. This process is repeated until the ray leaves the volume domain or an absorption event happens. 

High dynamic range light maps are employed to look up lighting information from the environment. 
Next event estimation is used to importance-sample rays towards the light source and potentially reduce the variance of the rendered image.
All Monte Carlo samples are accumulated and averaged in a floating-point accumulation buffer. A tone-mapping pass maps the accumulated result into the final lower dynamic range output buffer.
For fast image generation, we apply performance optimization methods such as empty-space skipping (based on the transfer function preset) and memory coherent scattering. The latter optimization ensures that rays of neighboring pixels are scattered in the same direction, thus ensuring optimized cache utilization.

\subsection{Compressed Differentiable 3D Gaussian Splatting}\label{sec:3dgs}


Differentiable 3DGS describes an objects by a set of 3D Gaussians 
\begin{equation}
    G(x) = \alpha e^{-\frac{1}{2}x^T\Sigma^{-1}x}.
\end{equation}
Each Gaussian is centered at $x\in\mathbb{R}^3$, and the covariance matrix $\Sigma \in\mathbb{R}^{3\times3}$ describes its orientation and shape. 
A Gaussian has an opacity $\alpha \in [0,1]$, and a view-dependent color that is represented by a set of spherical harmonics (SH) coefficients. 

The 2D projection of a 3D Gaussian is a 2D Gaussian with a covariance 
that is derived  
from the view transformation matrix and the Jacobian of the affine approximation of the projective transformation. 
The scene is rendered by projecting all Gaussian into the image plane in sorted order and blending their contributions. 


While Zwicker \etal~\cite{zwicker_ewa_2001} model a 3D scalar field via a set of 3D Gaussians so that the field can be reconstructed sufficiently well, Kerbl \etal~\cite{kerbl3Dgaussians} optimize the position, shape, opacity and SH coefficients of each 3D Gaussian so that their rendering matches a set of initial images of the object. The optimization is performed via differentiable rendering, by taking into account the changes in pixel color due to changes of the 3D Gaussian parameters. The optimization process removes some of the initially selected 3D Gaussians (if no contribution), adaptively splits Gaussians, and modifies their shapes and appearance attributes to minimize an image-based loss function. 


To further reduce the memory consumption of 3DGS, we utilize the compression proposed by Niedermayr~\etal~\cite{niedermayr2024compressed}.
It encodes SH coefficients and Gaussian shape parameters into compact codebooks via sensitivity-aware vector quantization, and then fine-tunes the parameters on the training images. Quantization-aware training \cite{rastegari_xnor-net_2016} is used to represent the scene parameters with fewer bits. We call this strategy High-Rate-compression (HR-compression). We also provide an option that uses only quantization-aware training to reduce all scene parameters but the Gaussians' positions to an 8-bit representation during optimization. We will subsequently call this strategy High-Quality-compression (HQ-compression). 
Since CA requires the entire data set in focus, we can omit the scaling factor that is usually stored per Gaussian to represent scenes with objects in focus and surrounding background.    


\noindent{\textbf{Alpha Channel Reconstruction}} In contrast to classical novel view synthesis, where only RGB colors are reconstructed, in volume rendering applications also the per-pixel accumulated opacity (i.e., alpha) needs to be reconstructed for blending correctly over the background.
Therefore, we extend 3DGS to allow for differentiable rendering of images with alpha channel. 
We use a combination of per pixel L1 and SSIM Loss to accurately reconstruct the alpha channel of the volume rendered training images. As shown in \cref{sec:quality}, the reconstruction quality can be improved significantly in this way.

\section{Results and evaluation}

We analyze the performance, memory consumption and quality of the proposed pipeline for CA with a variety of high-resolution medical data sets showing different anatomical structures. Our 3DGS implementation is a modification of the code provided by Kerbl~\etal~\cite{kerbl3Dgaussians}. For compression and rendering, we use the settings described by Niedermayr~\etal~\cite{niedermayr2024compressed}.

\subsection{Data Sets}

The hierarchical phase-contrast tomography (HiP-CT) data was acquired at the European Synchrotron Radiation Facility (ESRF) in the context of the Human Organ Atlas project\cite{Walsh2021}.

\textit{Kidney} is a HiP-CT scan from beamline 5 of the complete left kidney from body donor LADAF-2020-27 downsampled to 50.16 \textmu m resolution ($1510 \times 1706 \times 1415$ voxels in size) and quantized to 8~bit precision.

\textit{Brain} is a HiP-CT scan from beamline 18 of the complete brain of body donor LADAF-2021-17 downsampled for rendering to 46.84 \textmu m resolution ($3224 \times 3224 \times 3585$ voxels in size) and quantized to 8~bit precision. While the kidney data set is publically available, the brain data has not been published yet.

\textit{Body} is a human CT angiography scan at resolution $317 \times 317 \times 835$ from the collection by Wasserthal~\cite{jakob_wasserthal_2023_10047292}, image id \emph{s0287}. The data set contains some semi-transparent material showing significant differences under directional lighting. We render it under complex lighting conditions to challenge 3DGS's reconstruction capabilities.

\begin{figure}[t!]
\centering
\includegraphics[width=\linewidth]{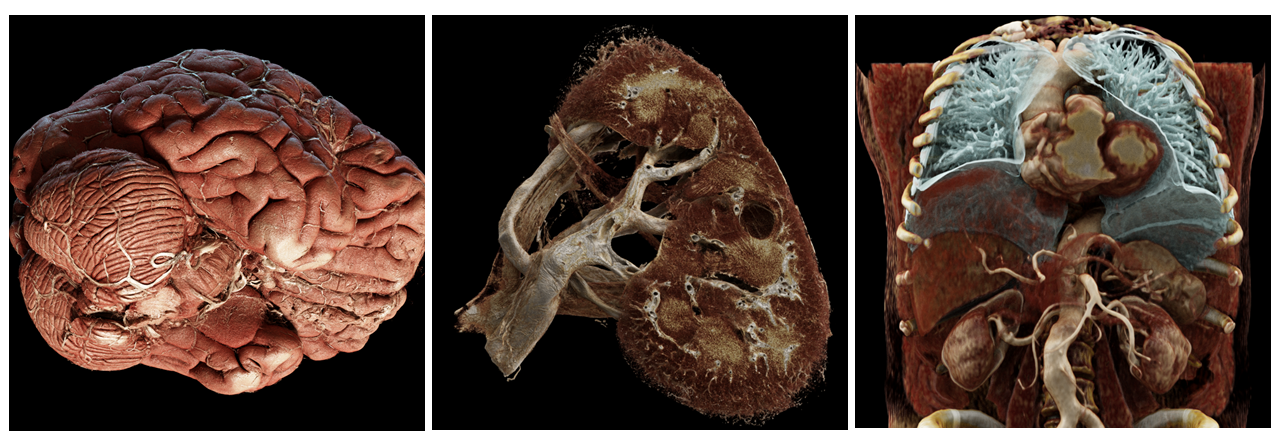}
\caption{From left to right, \textit{Brain} ($3224 \times 3224 \times 3585$), \textit{Kidney} ($1510 \times 1706 \times 1415$) and \textit{Body} ($317 \times 317 \times 835$). All images rendered at full HD with HR-compressed 3D Gaussian splatting using $<$ 70 MB per data set at $>$ 30 frames per second.}
\label{fig:datasets}
\end{figure}

We show all data sets in \cref{fig:datasets}, and provide an interactive online demonstration at \url{https://anonymous-demo-user.github.io/cinematic-3dgs/}. For each data set, between one and three  presets have been used, including segmentations, transfer function and lighting conditions. 3DGS optimization has been performed on training images of resolution $2048\times2048$.

\subsection{Preprocessing}

With a GPU providing sufficient RAM, the initial images of all data sets can be generated with the publically available CA package, 
and using the built-in animation system to generate the views. We have used
a research version providing batch rendering support, running on an NVIDIA A100 GPU for \textit{Brain} and an NVIDIA RTX A5000 for \textit{Kidney} and \textit{Body}.
\begin{table}[b]
\caption{Memory requirements and preprocessing performance.} 
    \label{tab:ablation}
    \resizebox{\linewidth}{!}{
\begin{tabular}{llll|lll}
\toprule
 & \multicolumn{3}{c}{Path Tracing} & \multicolumn{3}{c}{3DGS (HR-Compression)} \\
 & Size & Time & Views & Size & Time & Gaussians \\
\midrule
Brain & 36.4 GB & 158 Min & 99 & 69 MB & 106 Min & 4.8 M \\ 
Kidney & 3.6 GB & 6 Min & 101 & 33 MB & 53 Min & 2.3 M \\ 
Body & 0.2 GB & 23 Min & 99 & 7 MB & 50 Min & 0.9 M \\ 
\bottomrule
\end{tabular}
}
\end{table}

\cref{tab:ablation} shows in columns \emph{Size} the size of each data set in GB compared to the size of the final Gaussian representation in MB, when compressed using HR-compression. 
Column \emph{Views} shows the number of training images used for differentiable Gaussian splatting optimization. Columns \emph{Time} show the times to render the initial images via path tracing, and the computation times for generating the compressed Gaussian representations. Note that $90\%$ of the latter time are required by the optimization to generate the 3D Gaussian representation, and only about $10\%$ are consumed by the compression.   
Column \emph{Gaussians} gives the number of 3D Gaussians in the final representation. 
As can be seen in columns \emph{Size}, the compressed Gaussian representation is so small that it can be downloaded over low-bandwidth channels and rendered on mobile devices equipped with mid- or even low-end GPUs.



\subsection{View Selection}\label{sec:eval-viewsel}

\begin{figure}[t]
    \centering
    \includegraphics[width=0.9\linewidth]{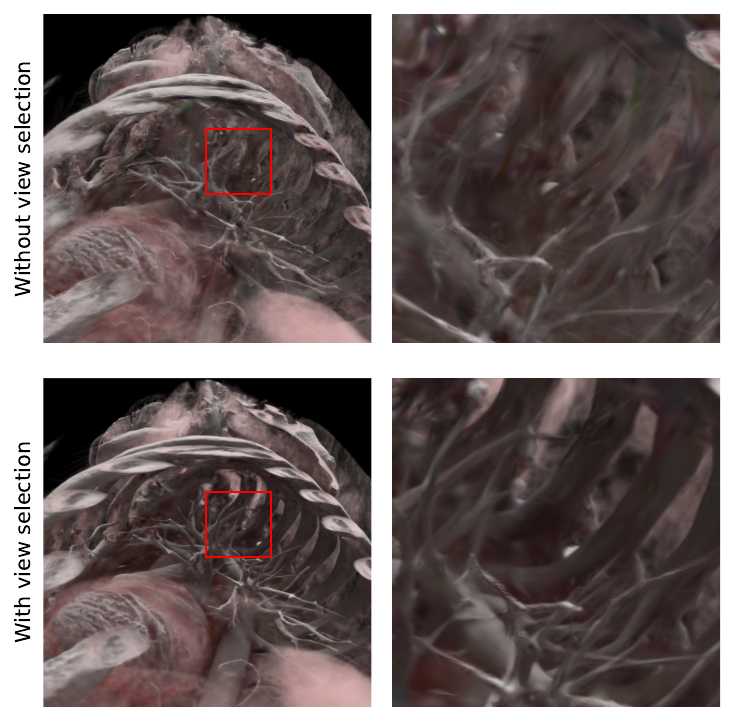}
    \caption{Automatic view selection generates camera poses covering unseen parts of the volume, like the inside of the rib cage. 
    }
    \label{fig:example-view-selection}
\end{figure}

Automatic view selection is demonstrated with Body, which exhibits a lot of structures that are not visible from cameras placed on an ellipsoid around the volume. 
As a baseline, we reconstruct the volume with images from 256 randomly placed cameras on the ellipsoid.
For comparison, we reduce this number to 128 and generate 128 additional cameras with the proposed view selection algorithm (see \cref{fig:example-view-selection}). As can be seen, overall improved reconstruction quality of parts not seen with random camera selection is achieved. 



\begin{figure*}[h!]
    \centering
\includegraphics[width=\linewidth]{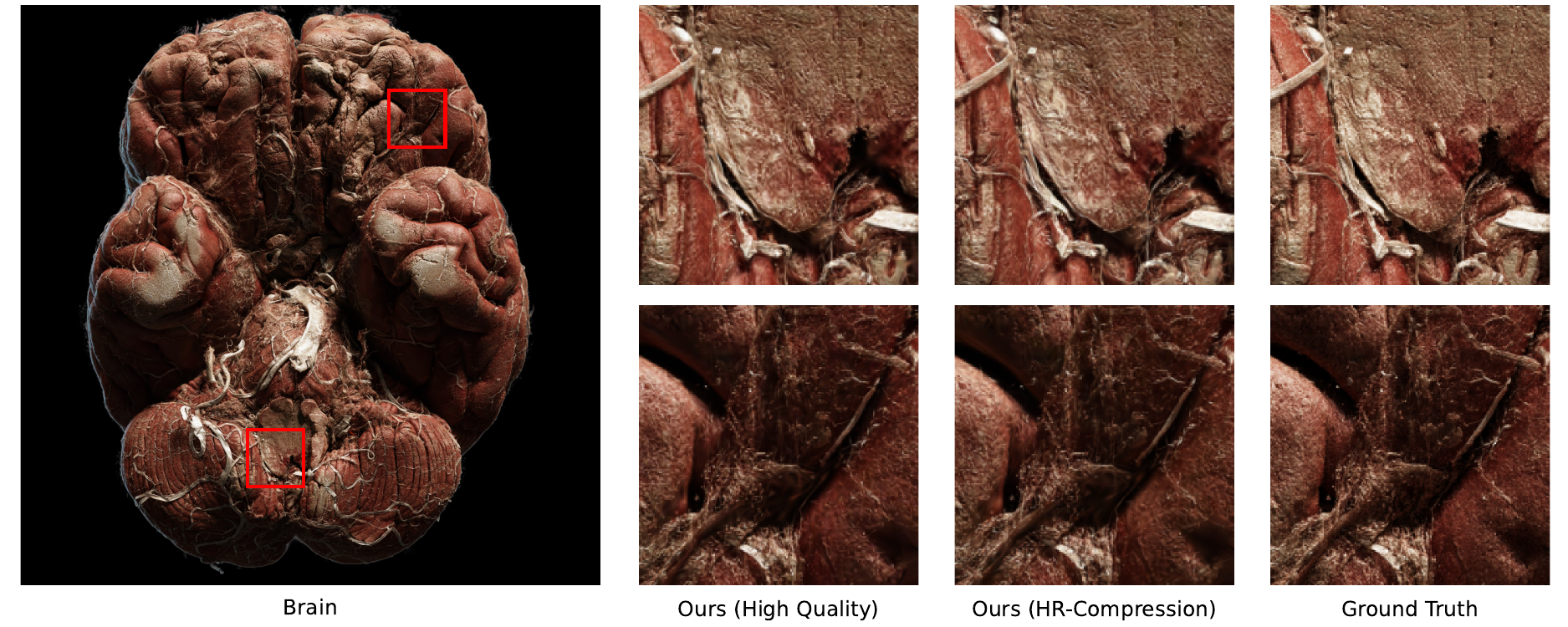}
\includegraphics[width=\linewidth]{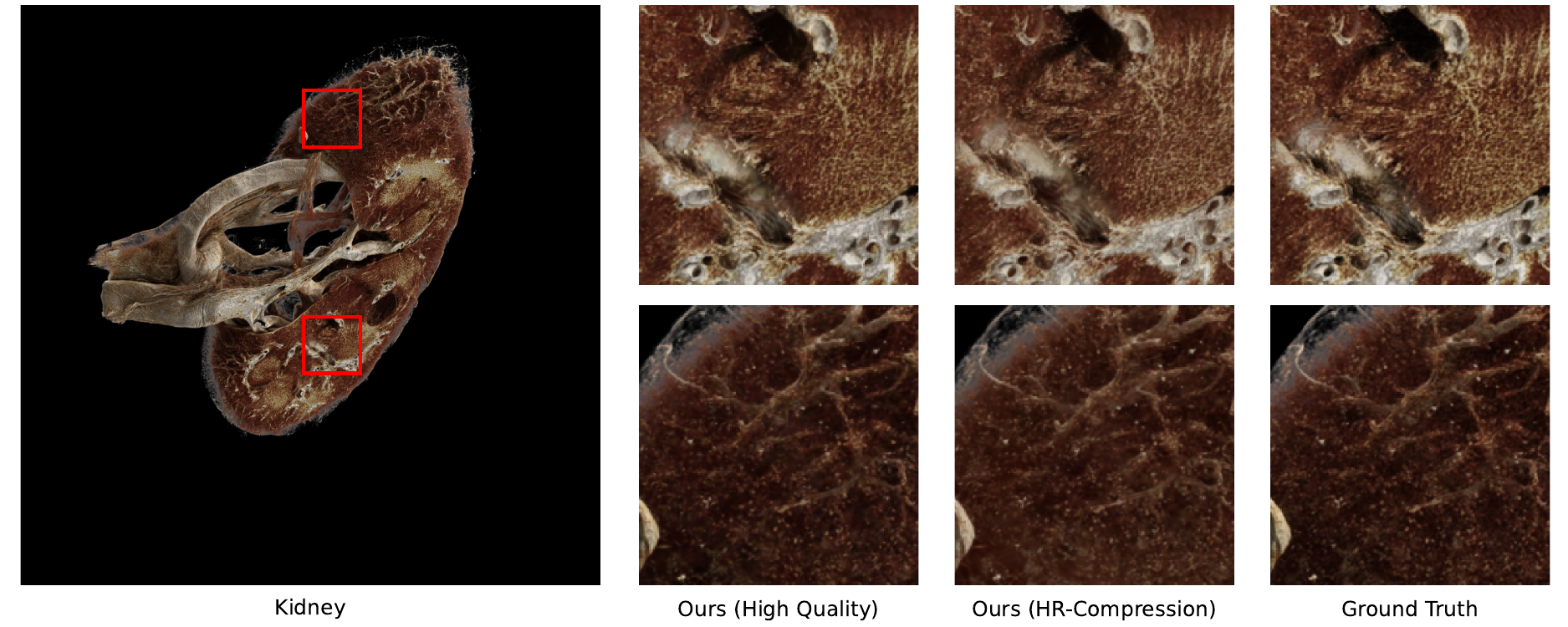}
\includegraphics[width=\linewidth]{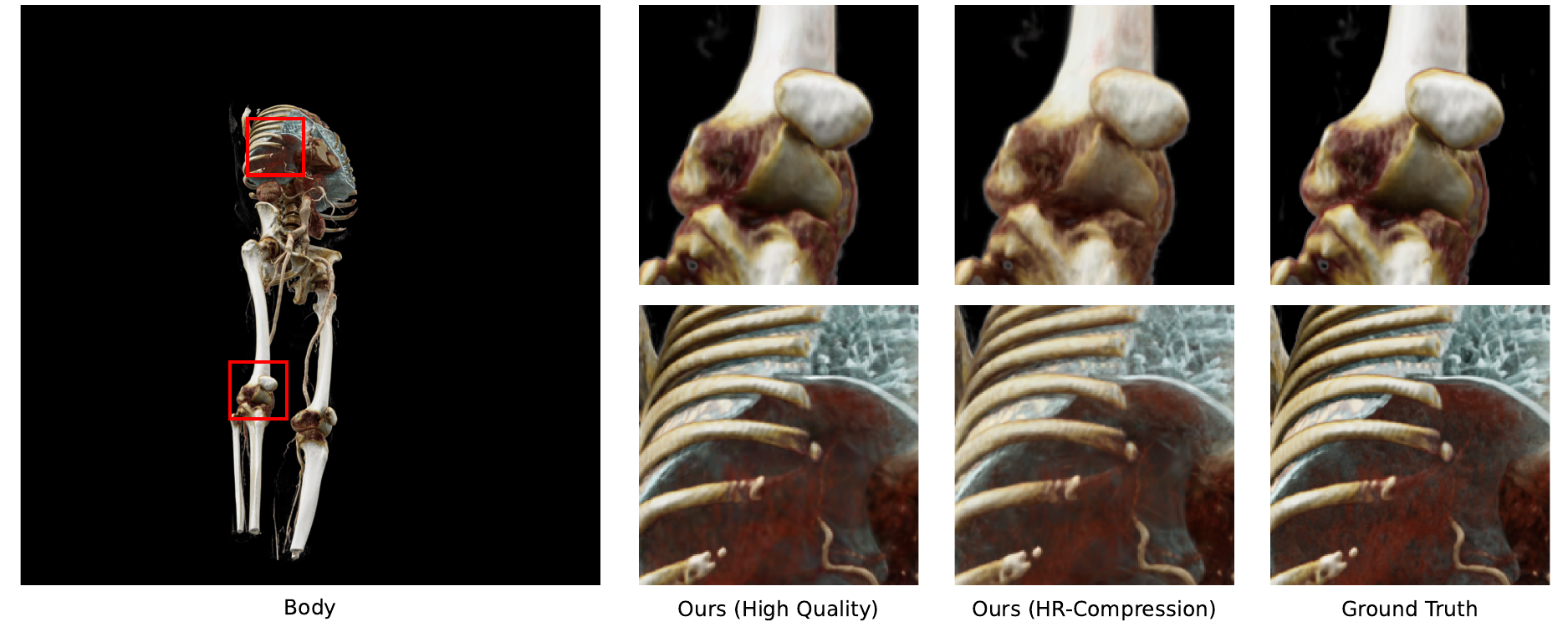}
    
    \caption{Quality comparison for HQ-compressed and HR-compressed Gaussian representations. All images are from the test set.}
    \label{fig:comparison-compression1}
\end{figure*}

\subsection{Quality Evaluation}\label{sec:quality}
\cref{fig:comparison-compression1} compares test images that have not been seen during 3DGS optimization to images rendered with HQ- and HR-compressed 3DGS. 
Close-up views reveal only subtle color shifts between path traced images and images generated via HR-compressed 3DGS. HQ-compression leads to an increase in memory by a factor of three, yet differences in image quality are further reduced and become so small that they are hardly noticeable by eye. 

\begin{figure}[t]
     \centering
     \includegraphics[width=\linewidth]{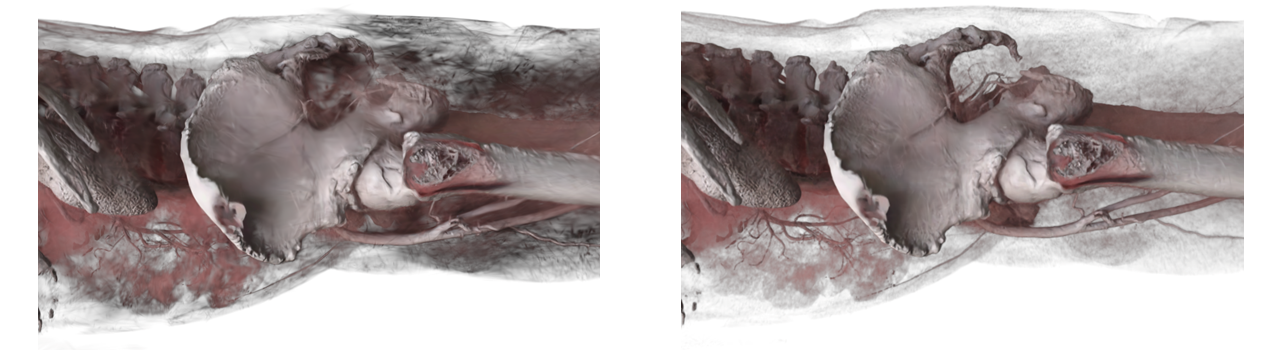}
    \caption{Color-only reconstruction (left) leads to reconstruction artifacts, which disappear when 3DGS is optimized for color and opacity (right).}
    \label{fig:alpha-comparison}
\end{figure}

Notably, significant losses in reconstruction quality are introduced when differentiable 3DGS optimizes only for RGB color (see \cref{fig:alpha-comparison} for an example). Extending 3DGS so that also opacity is considered in the optimization process improves greatly the reconstruction quality and removes unwanted artifacts caused by the background.

\cref{tab:results_2k} shows the average SSIM and PSNR between the test images and the novel views rendered with HR-compressed 3DGS, averaged over all presets. For PSNR and SSIM, only pixels which are not empty (alpha $>0$) in the rendered and ground truth image are considered. 
PSNR (Alpha) measures the PSNR for the alpha channel between rendered images and ground truths. 

\begin{table}[h]
\centering
\caption{Quantitative evaluation of HR-compression, 
}
\label{tab:results_2k}
\begin{tabular}{lccc}
\toprule
 & SSIM & PSNR & PSNR (Alpha) \\
Scene &  &  &  \\
\midrule
Brain & 0.72 & 23.23 & 34.09 \\
Kidney & 0.84 & 25.80 & 30.20 \\
Body & 0.87 & 26.90 & 29.57 \\
\bottomrule
\end{tabular}
\end{table}

\begin{figure}[h!]
    \centering
    \includegraphics[width=\linewidth]{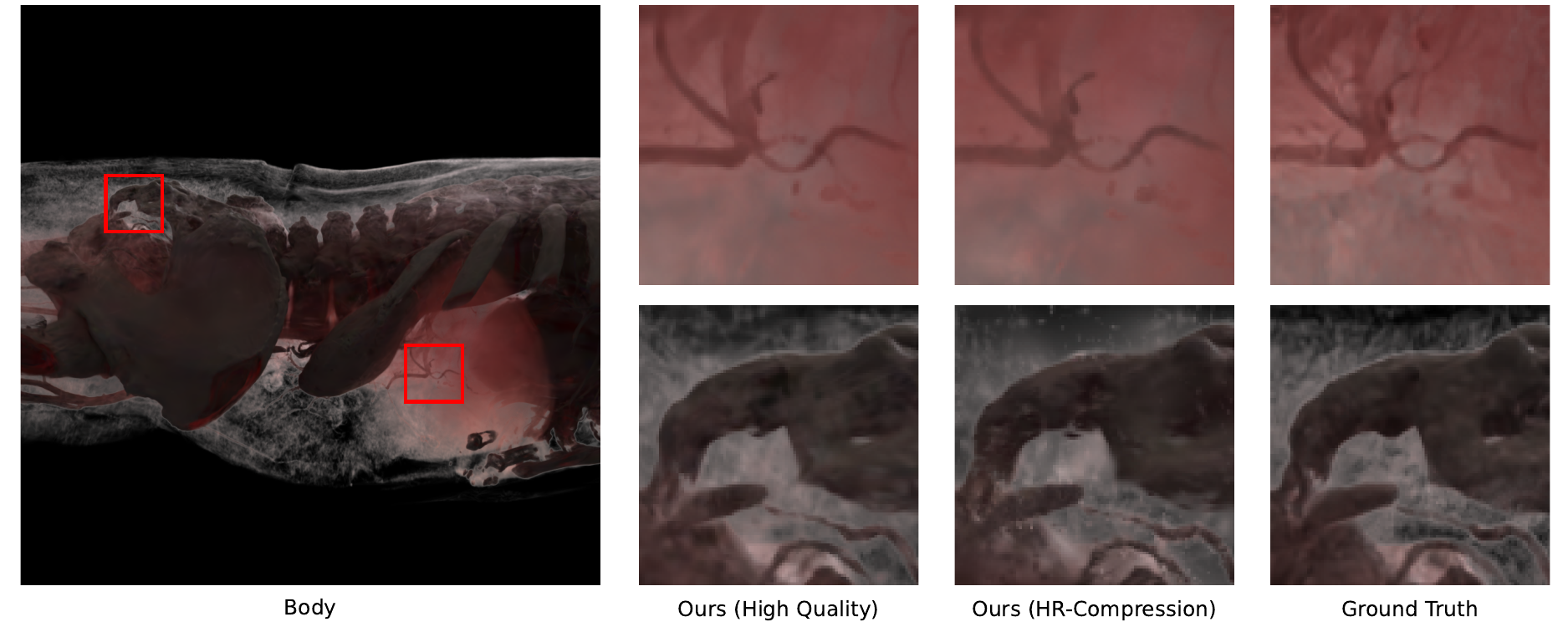}   
    \caption{Quality comparison for HQ-compressed and HR-compressed Gaussian representations of semi-transparent regions.}
    \label{fig:comparison-compression2}
\end{figure}

We further shed light on the capabilities of 3DGS to reconstruct semi-transparent regions in a data set. \textit{Body} is used with a preset so that certain tissue types in the data set become semi-transparent, see \cref{fig:comparison-compression2}.
While overall the novel view matches the test images fairly well, the closeup views show that some fine details are not reconstructed accurately, and especially the semi-transparent structures are blurred out. This effect increases with increasing depth complexity, since it becomes more and more difficult for 3DGS to represent all possible color and opacity distributions accurately. 

\begin{figure}[t]
     \centering
    \includegraphics[width=\linewidth]{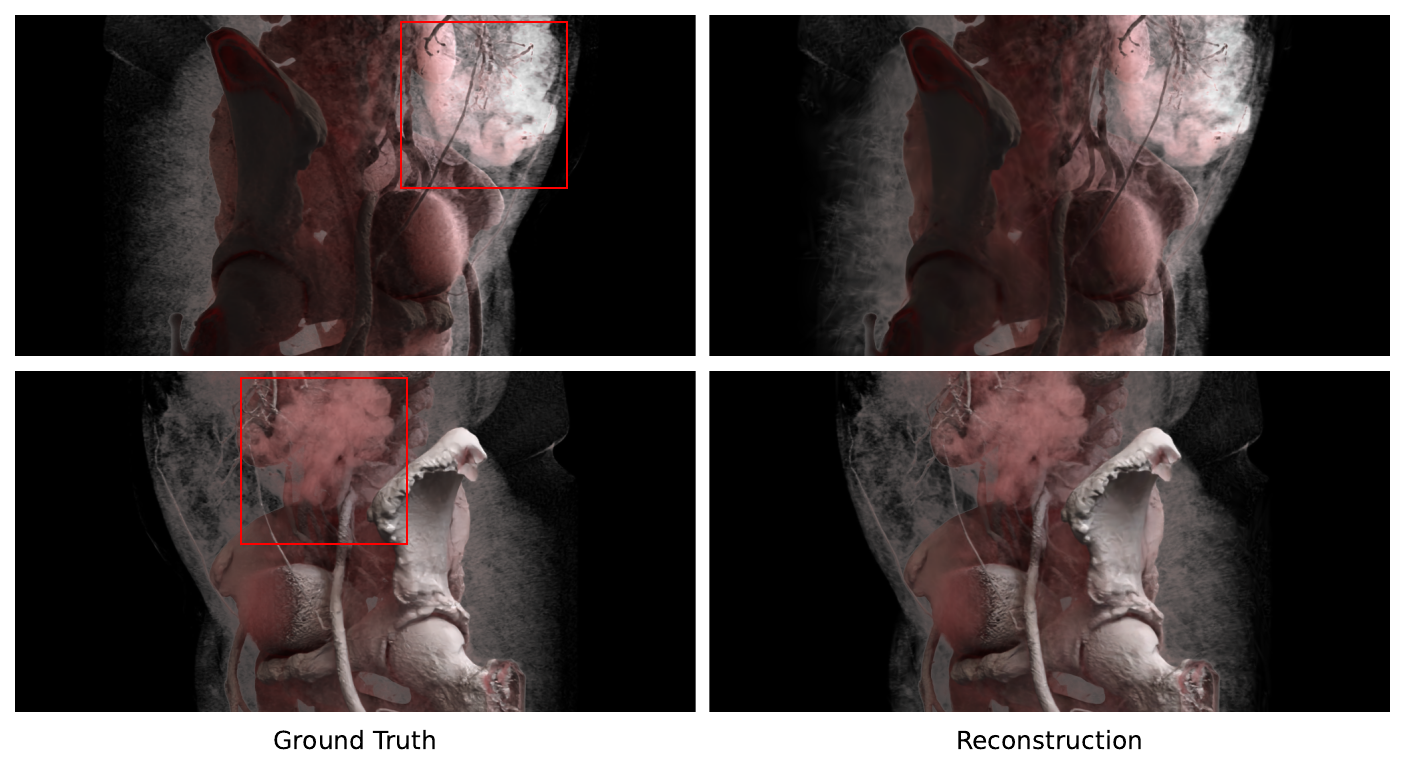}
    \caption{Complex view-dependent lighting effects are well preserved by compressed 3DGS, even for semi-transparent material. The tissue marked with a red box shows high-frequent color variation under different perspectives.}
    \label{fig:directional-color}
\end{figure}

When using a preset with strong directional lighting from the  environment map, one observes some high-frequency illumination variations especially in the volumetric regions. This makes it more difficult for 3DGS to accurately recover the tissue structures. Interestingly, \cref{fig:directional-color} demonstrates that the reconstruction works very well and does not show any severe reconstruction artefacts. At the same time, the semi-transparent regions are again blurred out to a certain extent. We believe that 3DGS has in particular problems with settings where the view rays accumulate matter over a long distance through semi-transparent, yet heterogeneous regions. In such situations, a subtle change of the camera pose can lead to strong changes of the per-pixel accumulated colors and opacities. Thus, 3DGS needs to optimize a significantly increased number of parameters, requiring far more Gaussians to accurately represent the data.  

\subsection{Rendering Performance}

The WebGPU implementation by Niedermayr \etal \cite{niedermayr2024compressed} is used for performance testing.
It enables rendering of compressed 3DGS up to $4\times$ faster than the renderer by Kerbl~\etal~\cite{kerbl3Dgaussians} and runs in a modern browser.


\cref{tab:render} shows that the rendering times even on an integrated iGPU 
is higher than 10 frames per second for the biggest data set \textit{Brain}.
On current mid- to high-end GPUs, 60 frames per second can be achieved for all data sets.
This makes the CA pipeline especially appealing for applications where stereoscopic rendering is required.
While low memory consumption facilitates efficient rendering on mobile devices, for instance, in mobile AR applications, high rendering performance is required to render two images (one for the left and one for right eye) at sufficient frame rates.
In a supplementary video we demonstrate rendering performance of roughly 5 to 20 frames per second on a mobile device with Qualcomm Adreno 740 GPU.


\begin{table}[h]
\resizebox{\linewidth}{!}{
\begin{tabular}{l|SSS}
\hline
                   &{Brain}   & {Body}  & {Kidney} \\ \hline
NVIDIA RTX 4070 TI Super   &  65 & 226  &  170 \\
NVIDIA RTX A5000      &  68 & 341  &  199 \\
AMD Ryzen™ 9 7900X iGPU    &  12 & 42  &  16 \\
\end{tabular}
}
\caption{Rendering performance at 2048x2048 resolution in frames per second, averaged over all training images and presets. For the iGPU a resolution of 1024x1024 is used.} 
\label{tab:render}
\end{table}




\section{Discussion and Outlook}
Our experiments show that compressed 3DGS enables interactive CA 
with extremely large data sets, 
by restricting to static presets. 
We believe that this limitation is acceptable for educational use since not more than a few presets are usually selected. Since the memory requirement of compressed 3DGS is so low, a separate Gaussian representation can be computed for each preset. 

\begin{figure}[h!]
    \centering
\includegraphics[width=\linewidth]{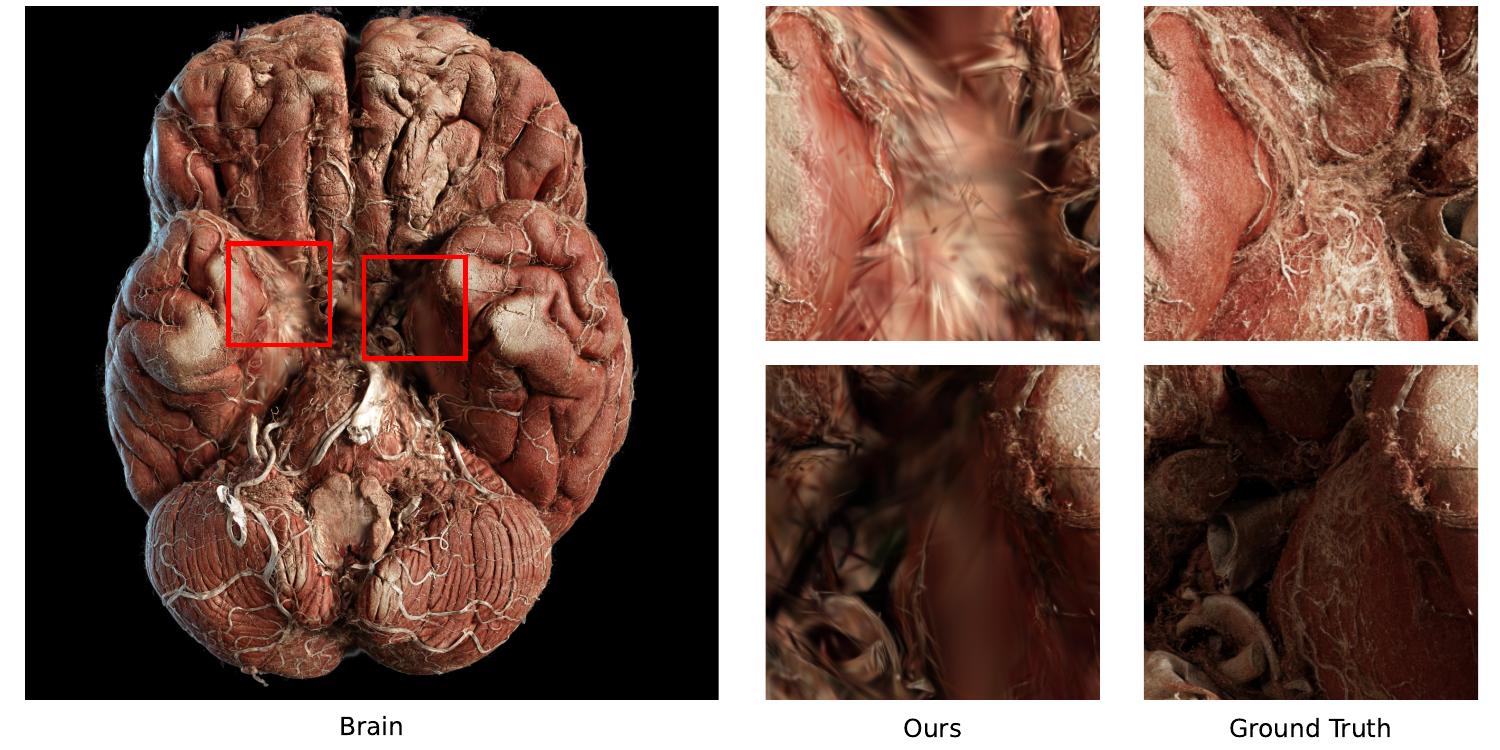}
    \caption{With a headlight, 3DGS faces problems in some places to accurately reconstruct structures with high-frequent changes in illumination.}
    \label{fig:scene-headlight}
\end{figure}

In all experiments
we have simulated static lighting conditions with an environment map that does not change relative to the object. 
Thus, the objects are seen under the same lighting condition in every view, resulting in rather smooth illumination when changing the camera pose.
This, however, changes when a headlight is used, and a point's illumination varies with varying camera pose (see \cref{fig:scene-headlight}). Notably, while in this situation most regions can be resolved very well by 3DGS, in some other regions the novel views show reconstruction artifacts. The strong variation of the reflected light under an illumination that changes in every image cannot be captured well by 3DGS.  
One approach we see to address this limitation is via re-lighting. By generating training images with optical material properties instead of illumination, it might be possible to better recover highly varying lighting conditions at runtime.

A useful component for volume exploration is an interactive clip plane, resulting in object points that were previously unseen to become visible.
One possibility to include clip planes is to restrict the plane movement to discrete steps and compute a separate Gaussian representation for each step. While this will significantly increase the memory requirements, we are confident that a fairly compact representation can be obtained by exploiting spatial coherence and progressively encoding the 3D Gaussians that appear and disappear when making subsequent steps.  



\section{Conclusion}

We have demonstrated the use of differentiable 3DGS for novel view synthesis from path traced images of high resolution medical data sets. We have shown that the 3D Gaussian representation can be compressed --- at hardly perceivable loss in image quality --- to a size that enables download and storage on even mobile devices. The Gaussian representation needs to be re-generated for every selected preset, yet even for many presets the overall memory is still significantly below the memory required by the data set. Computationally expensive path tracing can be avoided at rendering time, enabling fast display on mid- and even low-end devices. 

We have also pointed at current limitations of 3DGS for CA. As the most important ones we see the current absence of support for clip planes and the quality degradation when a headlight is used. We have sketched future research directions to address these limitations, and we are confident that improvements can be achieved. There is also a pressing need to handle time-varying data sets, since we see more and more scanning technologies that can accurately measure blood flow and deforming tissue. Tailoring 3DGS for interactively visualizing such dynamic processes is another important goal.    

Finally, we want to mention that besides CA we see in-situ visualization as another promising application of 3DGS. For data sets which are simulated on a supercomputer and are so large that they cannot be streamed out, images of the data set can be generated directly on the supercomputer and then streamed out to a system where novel view synthesis is performed. By using advanced implementations of 3DGS optimization, this might even become possible at rates enabling an explorative visual analysis.




\bibliographystyle{eg-alpha-doi} 
\bibliography{paper}

\clearpage


\twocolumn[
\centering
\Large
\vspace{0.5em}Supplementary Material \\
\vspace{1.0em}
]

\setcounter{subsection}{0}
\renewcommand\thesubsection{\Alph{subsection}}

\subsection{3DGS Optimizations}\label{sec:renderer}

\subsubsection{Volume Guided Initialization}

When using 3DGS, an initial set of 3D Gaussian kernels is first selected. These Gaussians are then removed, split or re-positioned, and the shape and appearance of the Gaussian kernels is optimized. Kerbl \etal~\cite{kerbl3Dgaussians} obtain the initial positions of the 3D Gaussians from the given images with structure from motion, or with random initialization where Gaussians are randomly positioned in the scene. For volume rendering, we randomly place Gaussians within the volume bounding box and set their initial color to grey. All other parameters are initialized as proposed by Kerbl \etal~\cite{kerbl3Dgaussians}.


Since in Cinematic Anatomy the 3D object and presets are known, an interesting question is whether the optimization process can be accelerated by initially placing Gaussians at locations were they will end up anyway. Thus, we initially position one Gaussian at every non-empty voxel in a low resolution version of the volume, and set the Gaussians' initial colors and opacities via the transfer function. Regions that are under-sampled by the initial sampling will be nevertheless represented by Gaussians due to adaptive splitting and relocation during optimization.   

\begin{figure}[h]
    \centering
    \includegraphics[width=\linewidth]{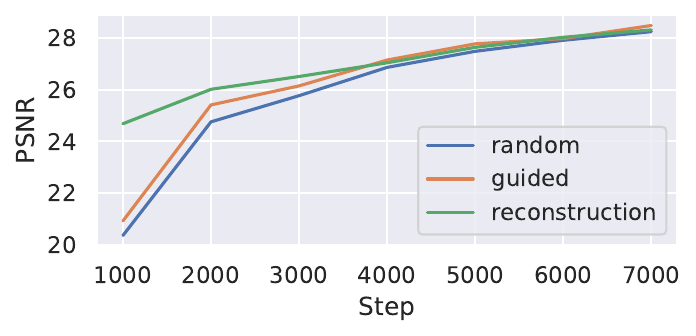}
    \caption{Test image reconstruction using different initialization schemes. Experiments were performed with \textit{Body}.}
    \label{fig:init}
\end{figure}

In \cref{fig:init} we exemplarily compare the effectiveness of the different initialization schemes based on optimization convergence for one of our test data sets. An initialization with the Gaussians' positions and colors from a previous reconstruction is used as gold standard. As can be seen, while all initialization techniques reach the same level of fidelity, volume-guided initialization does so with less iteration steps. However, it is fair to say that in all of our experiments the performance improvements were overall not significant, so that we decided to use random initialization in all tests.

\begin{table*}[hb]
    \centering
    \def\arraystretch{1.5}
\begin{tabular}{llllllrrrrr}
\toprule
 &  &  & Duration & Size & Points & Train Images & Test Images & SSIM & PSNR & PSNR (Alpha) \\
Scene & Preset & Resolution &  &  &  &  &  &  &  &  \\
\midrule
\multirow[t]{2}{*}{Brain} & 1 & 2k & 108 Min & 170 MB & 5.1 M & 87 & 12 & 0.64 & 20.73 & 32.76 \\
\cline{2-11}
 & 2 & 2k & 88 Min & 156 MB & 5.0 M & 87 & 12 & 0.81 & 25.78 & 35.86 \\
\cline{1-11} \cline{2-11}
\multirow[t]{2}{*}{Kidney} & 1 & 2k & 50 Min & 73 MB & 2.7 M & 91 & 13 & 0.81 & 23.93 & 27.70 \\
\cline{2-11}
 & 2 & 2k & 49 Min & 60 MB & 2.2 M & 87 & 12 & 0.87 & 27.89 & 32.85 \\
\cline{1-11} \cline{2-11}
\multirow[t]{4}{*}{Body} & 1 & 2k & 58 Min & 30 MB & 1.0 M & 87 & 12 & 0.89 & 28.93 & 31.54 \\
\cline{2-11}
 & 2 & 2k & 43 Min & 28 MB & 1.0 M & 87 & 12 & 0.87 & 25.38 & 29.94 \\
\cline{2-11}
 & 3 & 2k & 43 Min & 30 MB & 1.0 M & 87 & 12 & 0.88 & 27.65 & 30.40 \\
\cline{2-11}
 & 4 & 1k & 20 Min & 55 MB & 1.9 M & 224 & 32 & 0.86 & 28.68 & 25.41 \\
\cline{1-11} \cline{2-11}
\end{tabular}

    \caption{High Quality reconstruction details for all scenes and presets.}
    \label{tab:HQ-numbers}
\end{table*}

\begin{table*}[hb]
    \centering
    \def\arraystretch{1.5}
\begin{tabular}{lllrrlrrr}
\toprule
 &  & & Size & Gaussians & SSIM & PSNR & PSNR (Alpha) \\
Scene & Preset & Resolution &  &  &  &  &  &  \\
\midrule
\multirow[t]{2}{*}{Brain} & 1 & 2k & 69 & 4.8 M & 0.63 & 20.67 & 32.48 \\
\cline{2-9}
 & 2 & 2k & 69 & 4.8 M & 0.81 & 25.79 & 35.70 \\
\cline{1-9} \cline{2-9}
\multirow[t]{2}{*}{Kidney} & 1 & 2k & 37 & 2.6 M & 0.81 & 23.91 & 27.81 \\
\cline{2-9}
 & 2 & 2k & 30 & 2.1 M & 0.87 & 27.70 & 32.58 \\
\cline{1-9} \cline{2-9}
\multirow[t]{4}{*}{Body} & 1 & 2k & 7 & 1.0 M & 0.88 & 28.51 & 30.60 \\
\cline{2-9}
 & 2 & 2k & 7 & 0.9 M & 0.86 & 25.03 & 29.08 \\
\cline{2-9}
 & 3 & 2k & 7 & 1.0 M & 0.87 & 27.17 & 29.03 \\
\cline{2-9}
 & 4 & 1k  & 22 & 1.5 M & 0.81 & 26.13 & 24.46 \\
\cline{1-9} \cline{2-9}
\end{tabular}
\caption{High Range compression details for all scenes and presets}
\label{tab:HR-numbers}
\end{table*}

\subsubsection{Mip Splatting}

Scenes rendered with 3DGS can show severe artifacts when novel camera perspectives diverge from those the 3D Gaussian representation was optimized for. 
Yu~\etal~\cite{Yu2023MipSplatting} name the following two reasons for this behavior: Firstly, the 3D Gaussian representation exhibits frequencies that are too high to be faithfully reconstructed by the used sampling rate. Secondly, during splat-based rendering, a 2D dilation filter is applied that causes artefacts when zooming out and 2D splats become too small.

The problem is mitigated by introducing a 3D smoothing (i.e., low-pass) filter which constrains the size of the 3D Gaussians based on the maximal sampling frequency induced by the input views. A 2D Mip filter is applied in image space to avoid under-sampling. 
We observe that this extension to 3DGS significantly improves the fidelity of the reconstructed volumes for varying zoom levels.


\subsection{Performance Statistics and Further Results}
\begin{figure}[h!]
    \centering
\includegraphics[width=\linewidth]{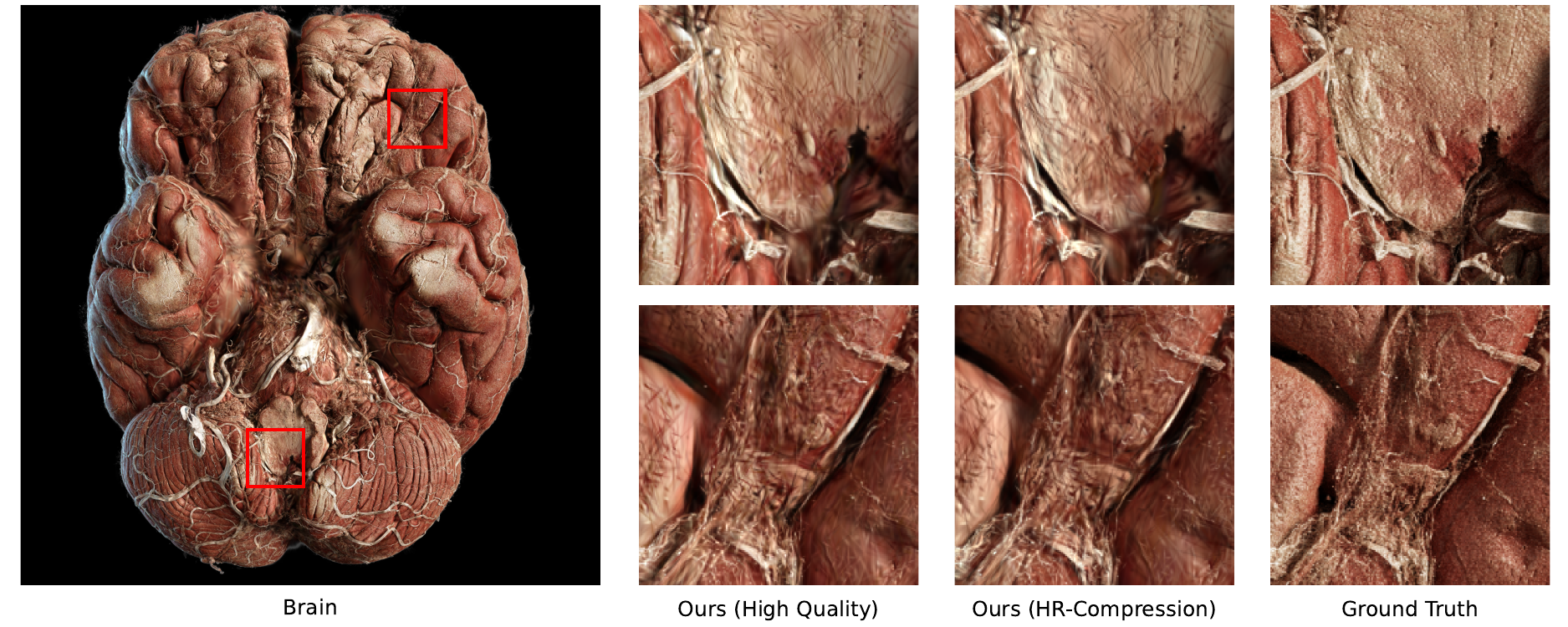}
\includegraphics[width=\linewidth]{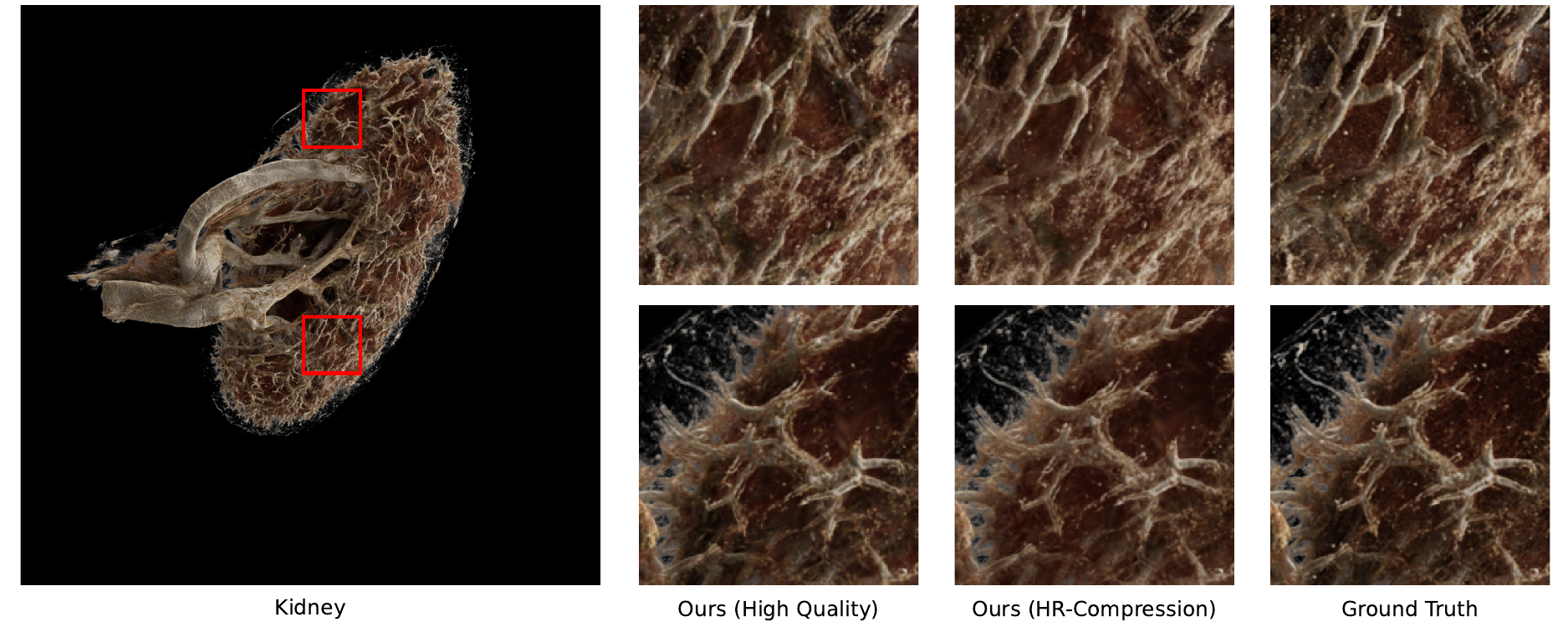}
\includegraphics[width=\linewidth]{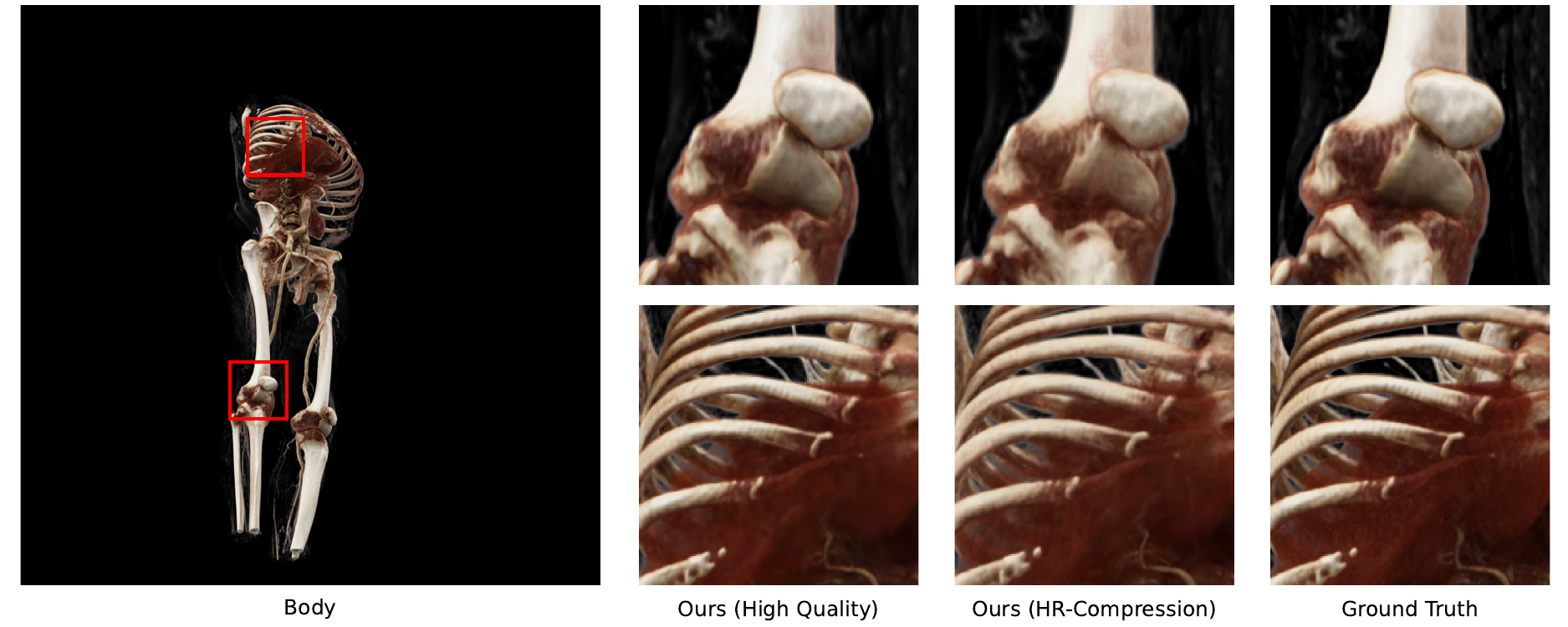}
\includegraphics[width=\linewidth]{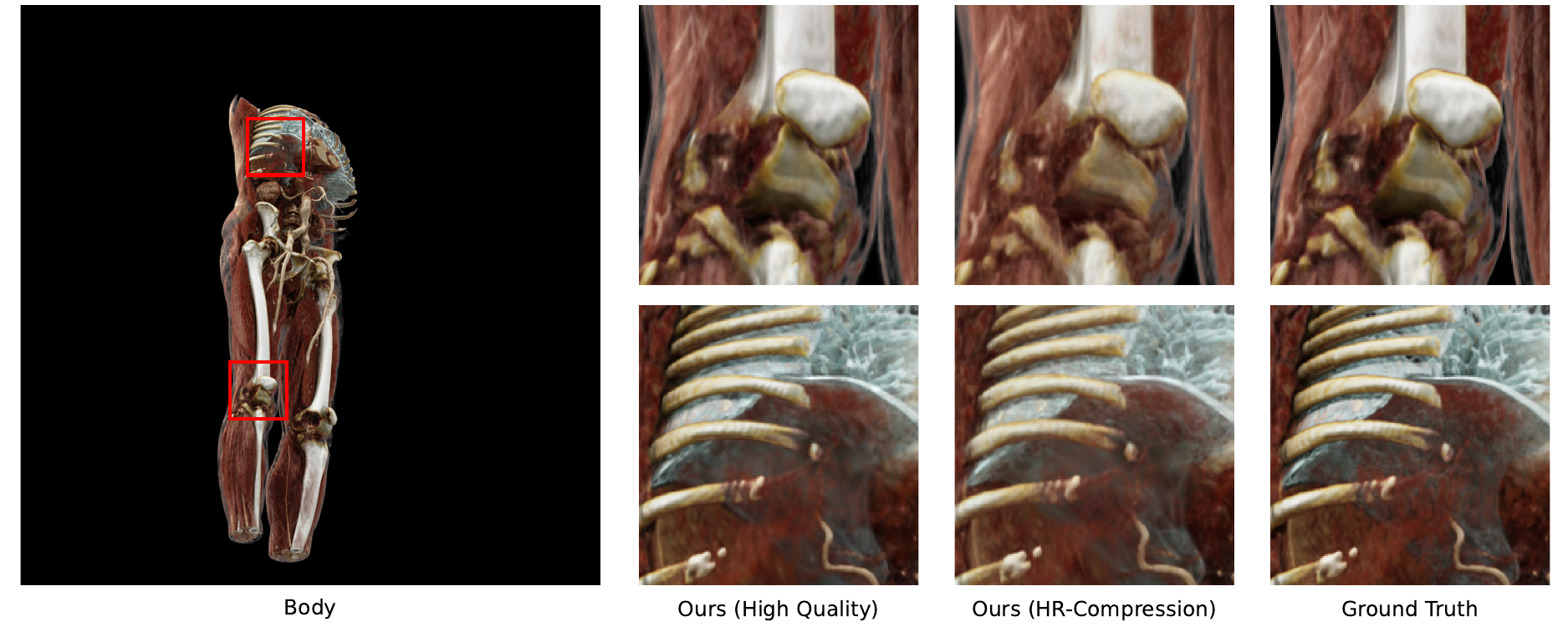}
\includegraphics[width=\linewidth]{figs/comparisons_vq/whole_body_preset4_4.pdf}
    
    \caption{Quality comparison for HQ-compressed and HR-compressed Gaussian representations. All images are from the test set.}
    \label{fig:comparison-compression-all}
\end{figure}

In \cref{tab:HQ-numbers} and \cref{tab:HR-numbers}, we provide detailed statistics using HR- and HQ-compression for all used presets. Additional qualitative results are shown in \cref{fig:comparison-compression-all}.

\subsection{View Selection}

\begin{figure}[h]
    \centering
    \includegraphics[width=\linewidth]{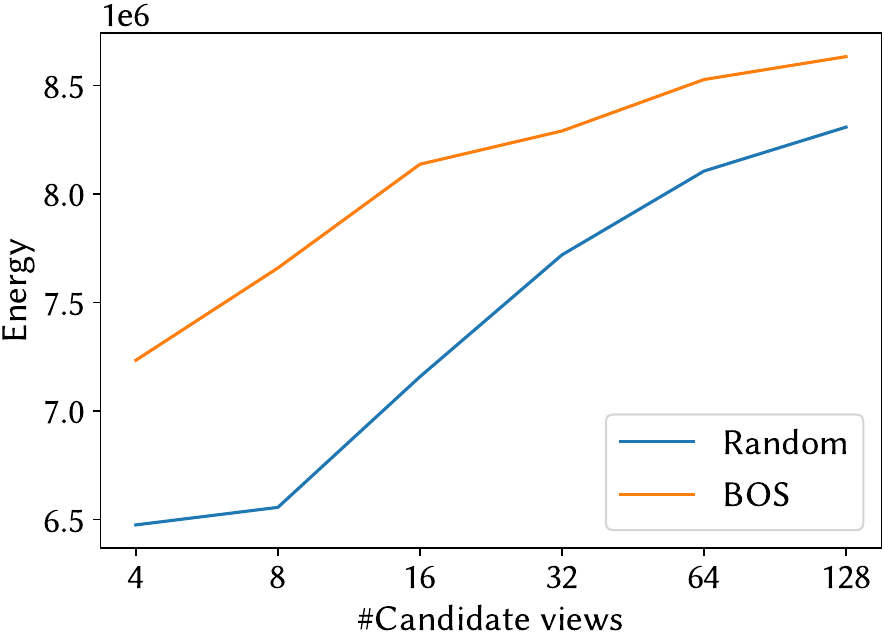}
    \caption{Achieved energy (the higher the better) when 32 camera poses are selected to generate test images of \textit{Body}. On the $x$-axis, the batch size of candidate poses in each iteration is shown. On the $y$-axis, the energy, as adapted from Kopanas and Drettakis~\cite{kopanas2023improving}, is shown.}
    \label{fig:bos-max}
\end{figure}

\begin{figure}[h]
    \centering
    \includegraphics[width=\linewidth]{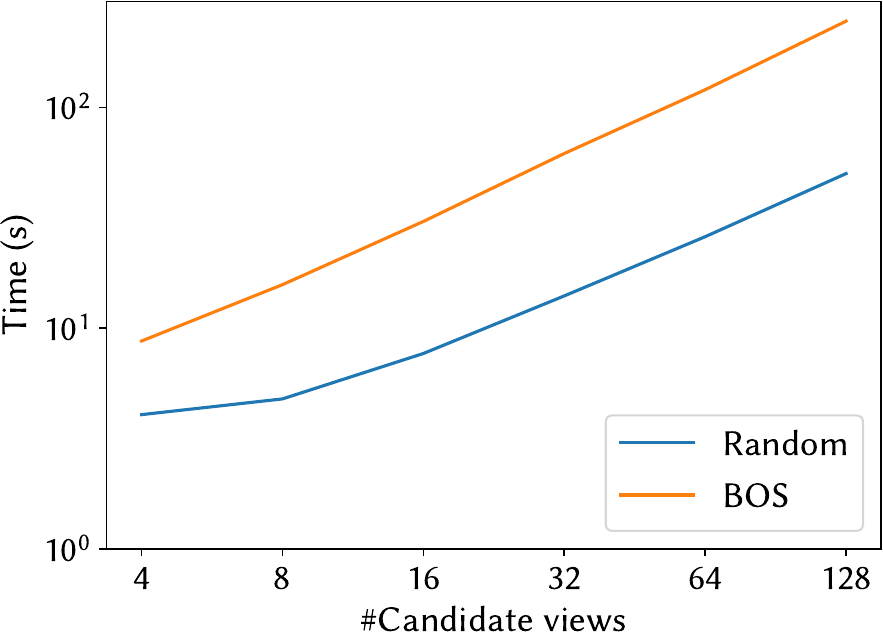}
    \caption{Run time for selecting 32 camera poses for \textit{Body}. On the $x$-axis, the batch size of candidate poses in each iteration is shown. On the $y$-axis, the run time is shown.}
    \label{fig:bos-time}
\end{figure}

In \cref{fig:bos-max}, we compare the convergence rate of BOS-based view selection to the approach of Kopanas and Drettakis~\cite{kopanas2023improving}, using preset 4 of \textit{Body}. The final energy term adapted from Kopanas and Drettakis~\cite{kopanas2023improving} is plotted after selecting a total of 32 cameras using a variable number of tested candidate views. \cref{fig:bos-time} shows the time it takes the view selection algorithms to select the 32 camera poses for these two strategies. BOS is able to find better maxima than pure random sampling of the candidate camera poses as described by Kopanas and Drettakis, albeit at a slightly higher computation time for the same number of generated candidate poses. The tests were run on a system with an AMD Ryzen 9 3900X 12-core (24-thread) CPU and an NVIDIA GeForce RTX 3090 GPU and averaged over 8 different random seeds. The BOS algorithm is run on the CPU, while the computation of the transmittance as described in the main manuscript is performed on the GPU using ray marching of a downscaled version of the volume.

\end{document}